\documentclass[11pt,a4paper]{article}

\usepackage[cp1251]{inputenc}
\usepackage[T2A]{fontenc}
\usepackage[english]{babel}

\usepackage[dvips]{graphicx}
\usepackage[dvips]{graphicx}
\usepackage{amsmath}
\usepackage{amssymb}
\usepackage{amsxtra}
\usepackage{amsthm}
\usepackage{latexsym}
\usepackage{array}
\usepackage{multirow}
\usepackage{hhline}
\usepackage[in]{fullpage}
\usepackage{color}
\newcommand {\ds}{\displaystyle}
\newcommand {\ri} {\mathrm{i}}

\begin{document}

\begin{center}

\Large{{\bf Reply to "Comment"\ by
A.\,V.~Tsiganov}}

\vspace{3mm}

P.\,E.~Ryabov

\vspace{3mm}

\small{\textit{Finance University under the
Government of the Russian Federation, Moscow,
Russia}

\textit{e-mail: orelryabov@mail.ru}}

\end{center}
\begin{flushright}
18.12.2011
\end{flushright}
\begin{abstract}
For the Goryachev case we obtain, in the
explicit form, the Abel-Jacobi equations with
the polynomial of degree six under the radical.
We choose the parameters of two families of
linear generators of a one sheet hyperboloid to
be the separation variables . These variables,
as well as the shifted separation variables in
the original work of S.~Kowalevski, do not
commute.
\end{abstract}

\vspace{5mm}

This reply is written as an answer to the paper \cite{tsig_archiv}, \cite{Tsig_rem_ND}.

\textbf{1.} In \cite{tsig_archiv}, \cite{Tsig_rem_ND} Tsiganov states that the
variables $u_1$, $u_2$ introduced in \cite{Ryab_Dan} are not the variables of separation
since they do not commute with respect to the initial Poisson bracket even on the zero-level
of the area integral. Let us write down all equations obtained in \cite{Ryab_Dan}, which were not shown by Tsiganov in "Comment".

Denote
\begin{equation*}\label{xx3_4}
\begin{array}{l}
  p_1(u)=2b + k -u^2, \qquad p_2(u)=2b -k +u^2, \qquad
  p_3(u)=(u-b)^2-f^2,\\
  p_{ij}=p_i(u_j), \qquad r_{ij}=\sqrt{p_{ij}} \qquad (i=1,2,3;\quad
  j=1,2).
\end{array}
\end{equation*}
and suppose that $u_{1,2}$ are the roots of the quadratic equation
\begin{equation}\label{xx3_2}
    z u^2-2b u+(2b \xi-k z)=0,\quad
z=\alpha_3^2,\quad
\xi=M_1^2+M_2^2+\frac{b}{\alpha_3^2}.
\end{equation}

\textrm{T h e o r e m  2. \cite{Ryab_Dan}}
\textit{The variables $u_1,u_2$ are separation
variables, and their evolution is described by
the Abel-Jacobi equations}
\begin{equation}\label{AJ}
\frac{du_1}{\sqrt{W(u_1)}}-\frac{du_2}{\sqrt{W(u_2)}}=0,\qquad
\frac{u_1 du_1}{\sqrt{W(u_1)}}-\frac{u_2
du_2}{\sqrt{W(u_2)}}=dt,
\end{equation}
\textit{where}
\begin{equation*}\label{xx3_14}
\begin{array}{l}
    W(u)=b^{-1} p_1(u) p_2(u) p_3(u)=\\[2mm]
    \phantom{W(u)}=b^{-1}(2b + k -u^2)(2b -k +u^2)[(u-b)^2-f^2].
\end{array}
\end{equation*}
\textit{Here the phase variables ${\boldsymbol
M},{\boldsymbol\alpha}$ are expressed
algebraically in terms of $u_1,u_2$ by the
formulas}
\begin{equation*}
\label{xx3_9}
\begin{array}{l}
M_1=\ri \ds {\frac{r_{21} r_{22}}{2 \sqrt{2b
(u_1+u_2)}}}, \quad M_2=-\ds{\frac{ r_{11}
r_{12}}{2 \sqrt{2b (u_1+u_2) }}},\\[5mm]
M_3=-\ds{\frac{\ri}{2 \sqrt{b}(u_1^2-u_2^2)}}( r_{12}  r_{22}  r_{31} +  r_{11}  r_{21}  r_{32}),\\[5mm]
\alpha_1=\ds{\frac{1}{2
\sqrt{b}(u_1^2-u_2^2)}}( r_{12}  r_{21}  r_{31}
+  r_{11}  r_{22}  r_{32}),\\[5mm]
\alpha_2=-\ds{\frac{\ri}{2
\sqrt{b}(u_1^2-u_2^2)}}( r_{11}  r_{22}  r_{31}
+  r_{12}  r_{21}  r_{32}), \\[5mm]
\alpha_3=\ds{\frac{ \sqrt{2 b}}{
\sqrt{u_1+u_2}} }.
\end{array}
\end{equation*}

This theorem is true. It can be proved by
direct substitution without using  any other
theory. Equations \eqref{AJ} could be obviously
written down as Kowalevski-type equations:
\begin{equation*}\label{xx3_13}
    (u_1-u_2)\frac{d u_1}{dt} = \sqrt{W(u_1)}, \qquad (u_1-u_2)\frac{d u_2}{dt} = \sqrt{W(u_2)}.
\end{equation*}

For experts, it is not difficult to conclude
whether the variables in these equations are
\textit{separation variables} or \textit{not}.
In \cite{Ryab_Dan} neither Lie-Poisson brackets  nor Hamiltonian formalism are mentioned.  The only
notion of differential equation theory we use is the notion of a first integral.

\textbf{2.} In \cite{tsig_archiv}, \cite{Tsig_rem_ND} Tsiganov points out the following: "At  $b=0$ this system and the
corresponding variables of separation have been investigated by Chaplygin \cite{chap}". Further he writes: "In \cite{tsig_pomi} we proved
that the Chaplygin variables remain variables of separation for the Goryachev case at $b\neq 0$".

To clarify the situation, let us write down some formulas. The papers \cite{tsig_pomi} and \cite{tsig_math} are the same, and we will
refer to \cite{tsig_pomi} for definiteness.

For $b=0$ the Chaplygin separation variables
have the form \cite{chap}:
\begin{equation}\label{eq6}
\begin{array}{l}
\ds{s_{1,2}=\frac{M_1^2+M_2^2\pm
h}{c\alpha_3^2},}
\end{array}
\end{equation}
where
$h^2=(M_1^2-M_2^2+c\alpha_3^2)^2+4M_1^2M_2^2$.

The Chaplygin separation variables depend on the value of $h$, the square of which is the function
of dynamic variables and, \textit{at the same time}, (for $b=0$) the value of the first
integral \cite{chap}.

In \cite{tsig_pomi} the separation variables
$q_{1,2}$ are introduced as the roots of the
quadratic equation \cite[formula
(3.8)]{tsig_pomi}:
\begin{equation}\label{tsig_3_8}
\lambda^2-\left(\frac{M_1^2+M_2^2}{\alpha_3^2}+c\right)\lambda+\frac{cM_2^2}{\alpha_3^2}=0,
\end{equation}
where $c$ corresponds to the parameter $c_2$ in
the formula (3.8) in \cite{tsig_pomi}.

\textit{Functional} relation between variables
\eqref{eq6} and \eqref{tsig_3_8} could be
easily obtained as follows:
\begin{equation}\label{eq7}
q_k=\frac{c}{2}s_k+\frac{c}{2},\quad k=1,2.
\end{equation}

Section 3.1 in \cite{tsig_pomi} is devoted to
separation of variables. At the same time,
nothing is said about the direct relation
\eqref{eq7} between $q_{1,2}$ and the Chaplygin
variables. We emphasize that simple formula
\eqref{eq7} is first written here and can not
be found in \cite{tsig_pomi}. To the contrary,
at the end of Section 3.1 we read: "Remark 2.
At $c_4=0$, we have reproduced the Chaplygin
result...". This means that the relation to
Chaplygin's result in \cite{tsig_pomi} is
pointed out only for $c_4=0$, which corresponds
to $b=0$ in equation \eqref{xx3_2}, and,
therefore, the separation variables $q_{1,2}$
are presented by Tsiganov in \cite{tsig_pomi}
as new variables of separation. Thus, the
statement that the separation of variables in
Goryachev problem in terms of Chaplygin
variables is proved in \cite{tsig_pomi},
\textit{does not represent the fact}. Moreover,
in \cite{tsig_pomi} there are no references
even to the original paper \cite{gory} devoted
to this problem. In particular, it is shown in
\cite{ryab_archiv_1}, \cite{ryab_archiv_2},
\cite{Ryab_Dan} that the integral presented in
\cite{tsig_pomi} is not new and can be
expressed in terms of Goryachev integral. The
history of this question can be found in
\cite{Yehia_ref}.

The question, whether the variables \eqref{eq6} are the separation variables for the Goryachev
case, is the question of definition of separation variables. It could be eventually reduced to the question whether the function
\begin{equation*}
h^2=(M_1^2-M_2^2+c\alpha_3^2)^2+4M_1^2M_2^2
\end{equation*}
is the first integral for the Goryachev case ($b\ne 0$)? The answer is evident.

\textbf{3.} In \cite{tsig_archiv},
\cite{Tsig_rem_ND} Tsiganov states that "an
application of the geometric Kharlamov method
to  the Goryachev system yields noncommutative
"new variables of separation" instead of the
standard canonical variables of separation",
and, in addition, he writes: "It is a
remarkable well-known shift of auxiliary
variables $u_{1,2}$, which Kowalevski used in
\cite{Kowa} in order to get canonical variables
of separation $s_{1,2}$ in her case".

Let us turn to the original paper \cite{Kowa} by Kowalevski, the letter \cite{letter_Kowa} of Kowalevski to Mittag-Leffler, as well as to the
original papers \cite{Ketter} and \cite{Appelrot} by K\"{o}tter and Appelrot.

The system of the first integrals is as follows:
\begin{equation*}\label{kowa_1}
\begin{array}{l}
\ds{2(p^2+q^2)+r^2=2\gamma_1+6l_1,}\\[3mm]
\ds{2(p\gamma_1+q\gamma_2)+r\gamma_3=2l,}\\[3mm]
\ds{\gamma_1^2+\gamma_2^2+\gamma_3^2=1,}\\[3mm]
\ds{\{(p+iq)^2+\gamma_1+i\gamma_2\}\{(p-iq)^2+\gamma_1-i\gamma_2\}=k^2,}
\end{array}
\end{equation*}
where $l_1,l$ and $k$ are real constants of the
first integrals.

S.\,Kowalevski introduced the polynomials:
\begin{equation*}\label{kowa_2}
\begin{array}{l}
\ds{R(x_1)=-x_1^4+6l_1x_1^2+4lx_1+1-k^2,}\\[3mm]
\ds{R(x_2)=-x_2^4+6l_1x_2^2+4lx_2+1-k^2,}\\[3mm]
\ds{R(x_1,x_2)=-x_1^2x_2^2+6l_1x_1x_2+2l(x_1+x_2)+1-k^2.}
\end{array}
\end{equation*}
Here
\begin{equation*}\label{kowa_3}
\ds{x_1=p+iq,\quad x_2=p-iq.}
\end{equation*}

In the letter to Mittag-Leffler, the founder of
the journal "Acta Mathematica"{\rm ,}
S.\,Kowalevski introduced the variables
$\frac{1}{2}w_1, \frac{1}{2}w_2$, where
\begin{equation}\label{kowa_4}
\begin{array}{l}
\ds{w_1=\frac{R(x_1,x_2)-\sqrt{R(x_1)}\sqrt{R(x_2)}}{(x_1-x_2)^2},}\\[5mm]
\ds{w_2=\frac{R(x_1,x_2)+\sqrt{R(x_1)}\sqrt{R(x_2)}}{(x_1-x_2)^2},}
\end{array}
\end{equation}
which at that moment were not shifted. In terms of these variables the Abel-Jacobi equations
are written down \cite[p.~166]{letter_Kowa}.

In \cite[p.~69]{Appelrot} Appelrot writes:
"With this, Kowalevski made her investigation
approximately in the way I show below, though
in some moments I make known deviations from
her following the example of
F.\,K\"{o}tter...". And, further, Appelrot
\cite[p.~70]{Appelrot} points out the
following:

"These values $w$, more precisely, the values
$s_1$ and $s_2$ that are equal to
\begin{equation}\label{kowa_6}
\ds{s_1=w_1+3l_1\qquad\text{and}\qquad
s_2=w_2+3l_1}
\end{equation}
are treated as \textit{new variables in the
analysis, however, Kowalevski, following closer Weierstrass, denoted by $s$ exactly what I denote by \mbox{$\overline{s}$}}\mbox{\ "\ } (italics and the equations number are mine; in Appelrot's book these equations have
the number $(10)$ in Page~70). In fact, the variables \eqref{kowa_6} were introduced by K\"{o}tter \cite{Ketter}.

Thus, the variables introduced by
S.\,Kowalevski \cite[formulas~(9),
p.~188]{Kowa}, are denoted in \cite{Appelrot}
by $\overline{s}$:
\begin{equation}\label{kowa_7}
\begin{array}{l}
\ds{\overline{s}_1=\frac{R(x_1,x_2)-\sqrt{R(x_1)}\sqrt{R(x_2)}}{2(x_1-x_2)^2}+\frac{1}{2}l_1,}\\[5mm]
\ds{\overline{s}_2=\frac{R(x_1,x_2)+\sqrt{R(x_1)}\sqrt{R(x_2)}}{2(x_1-x_2)^2}+\frac{1}{2}l_1.}
\end{array}
\end{equation}
The relation between Kowalevski variables
\eqref{kowa_7} and the variables \eqref{kowa_4}
and \eqref{kowa_6} can be also found in
\cite[p.~72]{Appelrot}:
\begin{equation*}
\ds{2\overline{s}+2l_1=s=w+3l_1.}
\end{equation*}

It is well known \cite{Veselov}, \cite{Komarov} that the variables $w_1$, $w_2$ do not commute
\begin{equation}\label{kowa_11}
\{w_1,w_2\}\ne 0,
\end{equation}
but the variables \eqref{kowa_6} do commute. They are used in classic works by K\"{o}tter \cite{Ketter}, Appelrot \cite{Appelrot},
Zhukovsky \cite{Zhuk}, Golubev \cite{Golubev}, Ipatov \cite{Ipatov}, and also in modern works devoted to constructing <<action-angle>>
variables (the latter should be canonic by definition, therefore the commutation property is motivated in these studies). Thus,
\begin{equation*}\label{kowa_10}
 \{s_1,s_2\}\equiv 0.
 \end{equation*}
Then
\begin{equation}\label{kowa_12}
\{\overline{s}_1,
\overline{s}_2\}=\frac{1}{4}[\{w_1,w_2\}-\{L_1,w_1-w_2\}]
\end{equation}
and
\begin{equation}\label{kowa_13}
0=\{s_1,s_2\}=\{w_1,w_2\}-3\{L_1,w_1-w_2\}.
\end{equation}
Here $L_1=\frac{1}{3}H$ is the first integral
with the constant $l_1$ ($H$ is the Hamiltonian function or the energy integral). From \eqref{kowa_13} we derive
\begin{equation}\label{kowa_14}
\{L_1,w_1-w_2\}=\frac{1}{3}\{w_1,w_2\}.
\end{equation}
Substituting \eqref{kowa_14} into the
expression \eqref{kowa_12} and using property
\eqref{kowa_11}, which is given in
\cite{Veselov}, \cite{Komarov} as self-evident,
we obtain
\begin{equation*}\label{kowa_15}
\{\overline{s}_1,
\overline{s}_2\}=\frac{1}{6}\{w_1,w_2\}\ne 0.
\end{equation*}

Thus, in the original paper \cite{Kowa} both pairs of the introduced variables $\frac{1}{2}w_1$, $\frac{1}{2}w_2$ and
$\overline{s}_1, \overline{s}_2$ (the latter pair is shifted
from $\frac{1}{2}w_1$, $\frac{1}{2}w_2$ by the
value $\frac{1}{2}l_1$) \textit{do not
commute}. There are no other separation
variables in the original Kowalevski paper. The
fact that the Kowalevski separation variables
do not commute is also mentioned in
\cite[p.~187]{bormam1}.

\textbf{4.} Conclusion.
\begin{itemize}
\item In \cite{Ryab_Dan} the Abel-Jacobi
equations with the polynomial of degree six under the radical are obtained for the
Goryachev case in the explicit form. For the separation variables we choose the parameters of
two families of linear generators of a one sheet hyperboloid. Whether they should commute
or not depends on definitions. As can be seen from the original paper \cite{Kowa}, the
Kowalevski separation variables \eqref{kowa_7} \textit{do not commute}.

\item In \cite{tsig_pomi} and \cite{tsig_math}, the fact that the Chaplygin variables in the Goryachev case are
separation variables (in any sense) is not mentioned clearly.

\item In the papers by Tsiganov cited above
there are no any references to the original
paper by Goryachev \cite{gory} devoted to this
problem and to the variants of constructing
separation variables given by Borisov and
Mamaev \cite{bormam1}, \cite{bormam2}.
\end{itemize}

\end{document}